\documentclass[amsmath,aps,prd,showpacs,a4paper,10pt]{revtex4}

\begin{document}

\newcommand{\be}[1]{\begin{equation}\label{#1}}
\newcommand{\ee}{\end{equation}}
\newcommand{\bea}{\begin{eqnarray}}
\newcommand{\eea}{\end{eqnarray}}
\def\disp{\displaystyle}

\begin{titlepage}

\begin{flushright}
astro-ph/0509328
\end{flushright}

\title{\Large \bf Cosmological Evolution of Hessence Dark Energy and Avoidance of the Big Rip}

\author{Hao Wei}
\email{haowei@itp.ac.cn}
 \affiliation{Institute of Theoretical Physics, Chinese Academy of
 Sciences,\\
P.O. Box 2735, Beijing 100080, China \\
 Graduate School of the Chinese Academy of Sciences, Beijing 100039, China}
 \author{Rong-Gen Cai}
 \email{cairg@itp.ac.cn}
 \affiliation{Institute of Theoretical Physics, Chinese Academy of
 Sciences,\\
P.O. Box 2735, Beijing 100080, China }

\begin{abstract}\vspace{1cm}
\centerline{\bf ABSTRACT}
Recently, many dark energy models whose equation-of-state parameter 
can cross the phantom divide $w_{de}=-1$ have been proposed. In a 
previous paper [Class. Quant. Grav. {\bf 22}, 3189 (2005); hep-th/0501160], 
we suggest such a model named hessence, in which a non-canonical 
complex scalar field plays the role of dark energy. In this work, the 
cosmological evolution of the hessence dark energy is investigated. We 
consider two cases: one is the hessnece field with an exponential potential, 
and the other is with a (inverse) power law potential. We separately 
investigate the dynamical system with four different interaction forms 
between hessence and background perfect fluid. It is found that the 
big rip never appears in the hessence model, even in the most general 
case, beyond particular potentials and interaction forms.
\end{abstract}


\maketitle

\end{titlepage}

\renewcommand{\baselinestretch}{1.6}



\section{Introduction}
Dark energy~\cite{r1} has been one of the focuses in modern
cosmology since the discovery of accelerated expansion of the
universe~\cite{r2,r3,r4,r5,r6,r40}. The simplest candidate of dark 
energy is a tiny positive cosmological constant. As an alternative to 
the cosmological constant, some dynamical scalar field models have
been proposed, such as quintessence~\cite{r7,r43}, 
phantom~\cite{r8,r9,r10,r41}, k-essence~\cite{r11} etc. In the
observational cosmology of dark energy, equation-of-state
parameter (EoS) $w_{de}\equiv p_{de}/\rho_{de}$ plays a central
role, where $p_{de}$ and $\rho_{de}$ are the pressure and energy
density of dark energy respectively. The most important difference
between cosmological constant and dynamical scalar fields is that
the EoS of the former is always a constant, $-1$,  while the EoS
of the latter can be variable during the evolution of the
universe.

Recently, by fitting the SNe Ia data, marginal evidence for
$w_{de}(z)<-1$ at $z<0.2$ has been found~\cite{r12}. In addition,
many best-fits of the present value of $w_{de}$ are less than $-1$
in various data fittings with different parameterizations
(see~\cite{r13} for a recent review). The present data seem to
slightly favor an evolving dark energy with $w_{de}$ being below
$-1$ around present epoch from $w_{de}>-1$ in the near
past~\cite{r14}. Obviously, the EoS $w_{de}$ cannot cross the
so-called phantom divide $w_{de}=-1$ for quintessence or phantom
alone. Some efforts~\cite{r15,r16,r17,r18,r19,r20,r21,
r22,r23,r24,r25,r26,r27,r28} have been made to build dark energy
model whose EoS can cross the phantom divide. Here we mention some
of them, such as the geometric approach (brane or
string)~\cite{r18}, holographic dark energy~\cite{r19,r20}, and
scalar-tensor theory (esp. non-minimal coupled scalar field to
gravity)~\cite{r27,r42}, etc.

Although some variants of k-essence~\cite{r11} look possible to
give a promising solution to cross the phantom divide, a no-go
theorem, shown in~\cite{r21}, shatters this kind of hopes: it is
impossible to cross the phantom divide $w_{de}=-1$, provided that
the following conditions are satisfied: (i) classical level, (ii)
general relativity is valid, (iii) single real scalar field, (iv)
arbitrary Lagrangian density $p\,(\varphi,X)$, where
$X\equiv\frac{1}{2}g^{\mu\nu}\partial_{\mu}\varphi\partial_{\nu}\varphi$
is the kinetic energy term, (v) $p\,(\varphi,X)$ is continuous
function and is differentiable enough, and (vi) the scalar field
is minimal coupled to gravity. Thus, to implement the transition
from $w_{de}>-1$ to $w_{de}<-1$ or vice versa, it is necessary to
give up at least one of conditions mentioned above.

Obviously, the simplest way to get around this no-go theorem is to
consider a model with  two real scalar fields, i.e. to break the
third condition. Feng, Wang and Zhang in \cite{r14} proposed a
so-called quintom model which is a hybrid of quintessence and
phantom (so the name quintom). Naively, one may consider a
Lagrangian density~\cite{r14,r22,r23,r24}
  \be{eq1} {\cal
L}_{quintom}=\frac{1}{2}\left(\partial_{\mu}\phi_1\right)^2-\frac{1}{2}\left(\partial_{\mu}
\phi_2\right)^2-V(\phi_1,\phi_2),
  \ee
  where $\phi_1$ and $\phi_2$
are two real scalar fields and play the roles of quintessence and
phantom respectively. Considering a spatially flat
Friedmann-Robertson-Walker (FRW) universe and assuming the scalar
fields $\phi_1$ and $\phi_2$ are homogeneous, one obtains the {\em
effective} pressure and energy density for the quintom
 \be{eq2}
p_{quintom}=\frac{1}{2}\dot{\phi}_{1}^2-\frac{1}{2}\dot{\phi}_{2}^2-V(\phi_1,\phi_2),~~~~~~~
\rho_{quintom}=\frac{1}{2}\dot{\phi}_{1}^2-\frac{1}{2}\dot{\phi}_{2}^2+V(\phi_1,\phi_2),
\ee
respectively. The corresponding {\em effective} EoS is given by 
\be{eq3}
w_{quintom}=\frac{\dot{\phi}_{1}^2-\dot{\phi}_{2}^2-2V(\phi_1,\phi_2)}{\dot{\phi}_{1}^2
-\dot{\phi}_{2}^2+2V(\phi_1,\phi_2)}.
 \ee
  It is easy to see that
$w_{quintom}\geq-1$ when $\dot{\phi}_{1}^2\geq\dot{\phi}_{2}^2$
while $w_{quintom}<-1$ when $\dot{\phi}_{1}^2<\dot{\phi}_{2}^2$.
The cosmological evolutions of the quintom model without direct
coupling between $\phi_1$ and $\phi_2$, i.e.
$V(\phi_1,\phi_2)=V_{\phi_1}+V_{\phi_2}\equiv
V_{\phi_{1}0}\exp(-\lambda_{\phi_1}\kappa\phi_1)
+V_{\phi_{2}0}\exp(-\lambda_{\phi_2}\kappa\phi_2),$ and with a
special interaction between $\phi_1$ and $\phi_2$, i.e.
$V(\phi_1,\phi_2)=V_{\phi_1}+V_{\phi_2}+V_{int}$ and
$V_{int}\sim\left(V_{\phi_1}V_{\phi_2}\right)^{1/2}$, were studied
by Guo {\it et al}~\cite{r22} and Zhang {\it et al}~\cite{r23},
respectively. They showed that the transition from
$w_{quintom}>-1$ to $w_{quintom}<-1$ or vice versa is possible in
this type of quintom model.

In~\cite{r28}, by a new view of quintom dark energy, we proposed a novel non-canonical
complex scalar field, which was named ``hessence'', to play the role of quintom. In the
hessence model, the phantom-like role is played by the so-called internal motion $\dot{\theta}$,
where $\theta$ is the internal degree of freedom of hessence. The transition from
$w_{h}>-1$ to $w_{h}<-1$ or vice versa is also possible in the hessence model~\cite{r28}.
We will briefly present the main points of hessence model in Sec.~\ref{sect2}.

The main aim of this work is to investigate the cosmological
evolution of hessence dark energy. We consider the hessence and
background perfect fluid as a dynamical system~\cite{r29}. By the
phase-space analysis, we find that some stable attractors can
exist, which are either scaling solutions or hessence-dominated
solutions with EoS ($w_h$ and $w_{eff}$) larger than or equal to
$-1$. {\em No} phantom-like late time attractors with EoS less
than $-1$  exist. This result is very different from the quintom 
model considered in~\cite{r22,r23}, where the phantom-dominated
solution is the {\em unique} late time attractor. If the universe
is attracted into the unique phantom-dominated attractor with EoS
less than $-1$, it will undergo a super-accelerated expansion
(i.e. $\dot{H}>0$) forever and the big rip is inevitable~\cite{r9}
(see also~\cite{r30,r16,r39}). On the contrary, in the hessence model, 
EoS less than $-1$ is transient. Eventually, it will come back to
quintessence-like attractors whose EoS is larger than $-1$ or
asymptotically to de Sitter attractor whose EoS is a constant
$-1$. Therefore, the big rip will not appear in the hessence
model.

The plan of this paper is as follows. In Sec.~\ref{sect2}, we will
briefly present the main points of the hessence model. In 
Sec.~\ref{sect3}, we give out the equations of the dynamical
system of hessence with/without interaction to background perfect
fluid for the most general case. That is, we leave the potential 
of hessence and the interaction form  undetermined. We will
investigate the dynamical system for the models with exponential
and (inverse) power law potentials in Sec.~\ref{sect4} and
Sec.~\ref{sect5}, respectively. In each case with different
potential, we consider four different interaction forms between
hessence and background perfect fluid. The first one corresponds
to the case without interaction, i.e. the interaction term is
zero. The other three forms are taken to be the most familiar
interaction ones considered in the literature. In all these cases,
we find that {\em no} phantom-like late time attractors with EoS
less than $-1$ exist. In Sec.~\ref{sect6}, we will show the big
rip will not appear in a general case beyond the above considered
cases. Finally, brief conclusion and discussions are given in
Sec.~\ref{sect7}.

We use the units $\hbar=c=1$, $\kappa^2\equiv 8\pi G$ and adopt the metric convention
as $(+,-,-,-)$ throughout this paper.


\section{\label{sect2} Hessence dark energy}
Following~\cite{r28}, we consider a non-canonical complex scalar
field as the dark energy, namely hessence,
 \be{eq4}
\Phi=\phi_1+i\phi_2, \ee with a Lagrangian density \be{eq5} {\cal
L}_h=\frac{1}{4}\left[\,(\partial_\mu \Phi)^2+(\partial_\mu
\Phi^\ast)^2\,\right]-U(\Phi^2+ \Phi^{\ast
2})=\frac{1}{2}\left[\,(\partial_\mu \phi)^2-\phi^2 (\partial_\mu
\theta)^2\,\right]-V(\phi),
 \ee
 where we have introduced two new
variables $(\phi,\theta)$ to describe the hessence, i.e.
 \be{eq6}
\phi_1=\phi\cosh\theta,~~~~~~~\phi_2=\phi\sinh\theta,
\ee
 which
are defined by
 \be{eq7}
\phi^2=\phi_{1}^2-\phi_{2}^2,~~~~~~~\coth\theta=\frac{\phi_1}{\phi_2}.
\ee
 Considering a spatially flat FRW universe with scale factor
$a(t)$ and assuming $\phi$ and $\theta$ are homogeneous, from
Eq.~(\ref{eq5}) we obtain the equations of motion for $\phi$ and
$\theta$,
 \bea
\ddot{\phi}+3H\dot{\phi}+\phi\dot{\theta}^2+V_{,\phi}=0, \label{eq8}\\
\phi^2\ddot{\theta}+(2\phi\dot{\phi}+3H\phi^2)\dot{\theta}=0,\label{eq9}
\eea
 where $H\equiv\dot{a}/a$ is the Hubble parameter, a dot and the subscript ``$_{,\phi}$'' denote the
derivatives with respect to cosmic time $t$ and $\phi$, respectively. The pressure and energy density
of the hessence are
\be{eq10}
p_h=\frac{1}{2}\left(\dot{\phi}^2-\phi^2\dot{\theta}^2\right)-V(\phi), ~~~~~~~
\rho_h=\frac{1}{2}\left(\dot{\phi}^2-\phi^2\dot{\theta}^2\right)+V(\phi),
\ee
respectively. Eq.~(\ref{eq9}) implies
\be{eq11}
Q=a^3 \phi^2\dot{\theta}=const.
\ee
which is associated with the total conserved charge within the physical volume due to the
internal symmetry~\cite{r28}. It turns out
\be{eq12}
\dot{\theta}=\frac{Q}{a^3 \phi^2}.
\ee
 Substituting this into
Eqs.~(\ref{eq8}) and (\ref{eq10}), we can recast them as
\be{eq13}
\ddot{\phi}+3H\dot{\phi}+\frac{Q^2}{a^6\phi^3}+V_{,\phi}=0,
\ee
\be{eq14}
p_h=\frac{1}{2}\dot{\phi}^2-\frac{Q^2}{2a^6\phi^2}-V(\phi),~~~~~~~
\rho_h=\frac{1}{2}\dot{\phi}^2-\frac{Q^2}{2a^6 \phi^2}+V(\phi).
\ee
It is worth noting that Eq.~(\ref{eq13}) is equivalent to the energy conservation equation
of hessence $\dot{\rho}_h+3H\left(\rho_h+p_h\right)=0$. The Friedmann equation and
Raychaudhuri equation are given by, respectively,
\be{eq15}
H^2=\frac{\kappa^2}{3}\left(\rho_h+\rho_m\right),
\ee
\be{eq16}
\dot{H}=-\frac{\kappa^2}{2}\left(\rho_h+\rho_m+p_h+p_m\right),
\ee
where $p_m$ and $\rho_m$ are the pressure and energy density of background matter,
respectively. The EoS of hessence $w_h\equiv p_h/\rho_h$. It is easy to see that $w_h\geq -1$
when $\dot{\phi}^2\geq Q^2/(a^6 \phi^2)$ while $w_h<-1$ when $\dot{\phi}^2< Q^2/(a^6 \phi^2)$.
The transition occurs when $\dot{\phi}^2=Q^2/(a^6 \phi^2)$.

There are other interesting points in the hessence model, such as
the avoidance of Q-ball formation, the novel possibility of
anti-dark energy, the possible relation between hessence and
Chaplygin gas, etc. We refer to the original paper~\cite{r28} for more details.


\section{\label{sect3} Dynamical system of hessence with/without interaction to background perfect fluid}

Now, we generalize the original hessence model~\cite{r28} to a
more extensive case. We consider a universe containing both
hessence dark energy and  background matter. The background matter
is described by a perfect fluid with  barotropic equation of state
\be{eq17}
 p_m=w_m\rho_m\equiv (\gamma-1)\rho_m,
 \ee
 where the
so-called barotropic index $\gamma$ is a constant and satisfies
$0<\gamma\leq 2$. In particular, $\gamma=1$ and $4/3$ correspond
to dust matter and radiation, respectively.

We assume the hessence and background matter interact through an
interaction term $C$, according to
 \bea
\dot{\rho}_h+3H\left(\rho_h+p_h\right)=-C,\label{eq18}\\
\dot{\rho}_m+3H\left(\rho_m+p_m\right)=C,\label{eq19}
 \eea
 which preserves the total energy conservation equation
$\dot{\rho}_{tot}+3H\left(\rho_{tot}+p_{tot}\right)=0$. Clearly,
$C=0$ corresponds to {\em no} interaction between hessence and
background matter. It is worth noting that Eq.~(\ref{eq13}) should
be changed when $C\not=0$, a new term due to $C$ will appear in
the right hand side. Since $\theta$ is the internal degree of
freedom~\cite{r28}, Eqs.~(\ref{eq9}), (\ref{eq11}) and (\ref{eq12}) still
hold. The hessence interacts to external matter only through
$\phi$. Thus, the equation of motion of $\phi$ should be changed
by the interaction.

Following~\cite{r31,r32,r33}, we introduce following dimensionless variables
\be{eq20}
x\equiv\frac{\kappa\dot{\phi}}{\sqrt{6}H},~~~y\equiv\frac{\kappa\sqrt{V}}{\sqrt{3}H},~~~
z\equiv\frac{\kappa\sqrt{\rho_m}}{\sqrt{3}H},~~~u\equiv\frac{\sqrt{6}}{\kappa\phi},~~~
v\equiv\frac{\kappa}{\sqrt{6}H}\frac{Q}{a^3 \phi}.
\ee
By the help of Eqs.~(\ref{eq14})--(\ref{eq16}), the evolution equations~(\ref{eq18}) and (\ref{eq19}) can
then be rewritten as a dynamical system:
\bea
&&\disp x^\prime=3x\left(x^2-v^2+\frac{\gamma}{2}z^2-1\right)-uv^2
-\frac{\kappa V_{,\phi}}{\sqrt{6}H^2}-C_1,\label{eq21}\\
&&\disp y^\prime=3y\left(x^2-v^2+\frac{\gamma}{2}z^2\right)
+\frac{\kappa}{2\sqrt{3}H}\frac{V_{,\phi}}{\sqrt{V}}\frac{\dot{\phi}}{H},\label{eq22}\\
&&\disp z^\prime=3z\left(x^2-v^2+\frac{\gamma}{2}z^2-\frac{\gamma}{2}\right)+C_2,\label{eq23}\\
&&\disp u^\prime=-xu^2,\label{eq24}\\
&&\disp
v^\prime=3v\left(x^2-v^2+\frac{\gamma}{2}z^2-1\right)-xuv,\label{eq25}
\eea where \be{eq26} C_1\equiv\frac{\kappa C}{\sqrt{6}H^2
\dot{\phi}}, ~~~~~~~ C_2\equiv\frac{\kappa
C}{2\sqrt{3}H^2\sqrt{\rho_m}},
\ee
  a prime denotes derivative
with respect to the so-called e-folding time ${\cal N}\equiv\ln
a$, and we have used
 \be{eq27}
-\frac{\dot{H}}{H^2}=3\left(x^2-v^2+\frac{\gamma}{2}z^2\right).
\ee The Friedmann constraint equation~(\ref{eq15}) becomes
\be{eq28}
 x^2+y^2+z^2-v^2=1.
  \ee
The fractional energy densities of hessence and background matter
are given by \be{eq29} \Omega_h=x^2+y^2-v^2,~~~~~~~\Omega_m=z^2,
\ee respectively. The EoS of hessence and the effective EoS of the
whole system are
 \be{eq30}
w_h=\frac{p_h}{\rho_h}=\frac{x^2-v^2-y^2}{x^2-v^2+y^2},~~~~~~~
w_{eff}=\frac{p_h+p_m}{\rho_h+\rho_m}=x^2-v^2-y^2+(\gamma-1)z^2,
\ee
 respectively. One can see that $w_h\geq -1$ as $x^2\geq v^2$, while $w_h<-1$
as $x^2<v^2$. Finally, it is worth noting that $y\geq 0$ and
$z\geq 0$ by definition, and in what follows, we only consider the
case of expanding universe with $H>0$.

It is easy to see that Eqs.~(\ref{eq21})--(\ref{eq25}) become an
autonomous system when the potential $V(\phi)$ is chosen to be an
exponential or (inverse) power law potential and the interaction
term $C$ is chosen to be a suitable form. Indeed, we will consider
the model with an exponential or (inverse) power law potential in
Sec.~\ref{sect4} and Sec.~\ref{sect5}, respectively. In each model
with different potential, we consider four cases with different
interaction forms between hessence and background perfect fluid.
The first case is the one without interaction, i.e. $C=0$. The
other three cases are taken as the most familiar interaction terms
extensively considered in the literature:
\begin{eqnarray*}
&{\rm Case~(I)} &C=0,\\
&{\rm Case~(II)} &C=\alpha\kappa\rho_m\dot{\phi},\\
&{\rm Case~(III)} &C=3\beta H\rho_{tot}=3\beta H\left(\rho_h+\rho_m\right),\\
&{\rm Case~(IV)} &C=3\eta H\rho_m,
\end{eqnarray*}
where $\alpha$, $\beta$ and $\eta$ are dimensionless constants. The 
interaction form Case~(II) arises from, for instance, string
theory or scalar-tensor theory (including Brans-Dicke
theory)~\cite{r33,r34,r35}. The interaction forms
Case~(III)~\cite{r36} and Case~(IV)~\cite{r37} are phenomenally
proposed to alleviate the coincidence problem.

In the next two sections, we  first obtain  the critical points
$(\bar{x},\bar{y},\bar{z},\bar{u},\bar{v})$ of the autonomous
system by imposing the conditions
$\bar{x}^\prime=\bar{y}^\prime=\bar{z}^\prime=\bar{u}^\prime=\bar{v}^\prime=0$.
Of course, they are subject to the Friedmann constraint, i.e.
$\bar{x}^2+\bar{y}^2+\bar{z}^2-\bar{v}^2=1$. We then  discuss the
existence and stability of these critical points. An attractor is
one of the stable critical points of the autonomous system.


\section{\label{sect4} Model with exponential potential}
In this section, we consider the hessence model with an
exponential potential
 \be{eq31}
 V(\phi)=V_0\, e^{-\lambda\kappa\phi},
\ee
 where $\lambda$ is a dimensionless
constant. Without loss of generality, we choose $\lambda$ to be
positive, since we can make it positive through field redefinition
$\phi\to -\phi$ if $\lambda$ is negative. In this case,
Eqs.~(\ref{eq21})--(\ref{eq25}) become
 \bea &&\disp
x^\prime=3x\left(x^2-v^2+\frac{\gamma}{2}z^2-1\right)-uv^2
+\sqrt{\frac{3}{2}}\lambda y^2-C_1,\label{eq32}\\
&&\disp y^\prime=3y\left(x^2-v^2+\frac{\gamma}{2}z^2\right)
-\sqrt{\frac{3}{2}}\lambda xy,\label{eq33}\\
&&\disp z^\prime=3z\left(x^2-v^2+\frac{\gamma}{2}z^2-\frac{\gamma}{2}\right)+C_2,\label{eq34}\\
&&\disp u^\prime=-xu^2,\label{eq35}\\
&&\disp v^\prime=3v\left(x^2-v^2+\frac{\gamma}{2}z^2-1\right)-xuv,\label{eq36}
\eea
  where $C_1$ and $C_2$ are defined by Eq.~(\ref{eq26}) and
depend on the interaction form $C$. In the following subsections,
we will consider different interaction forms $C$ mentioned in the
end of Sec.~\ref{sect3}.

To study the stability of the critical points of
Eqs.~(\ref{eq32})--(\ref{eq36}), we substitute linear
perturbations $x\to\bar{x}+\delta x$, $y\to\bar{y}+\delta y$,
$z\to\bar{z}+\delta z$, $u\to\bar{u}+\delta u$, and
$v\to\bar{v}+\delta v$ about the critical point
$(\bar{x},\bar{y},\bar{z},\bar{u},\bar{v})$ into
Eqs.~(\ref{eq32})--(\ref{eq36}) and linearize them. Note that
these critical points must satisfy the Friedmann constraint,
$\bar{y}\geq 0$, $\bar{z}\geq 0$ and requirement of
$\bar{x},\,\bar{y},\,\bar{z},\,\bar{u},\,\bar{v}$ all being real. 
Because of the Friedmann constraint~(\ref{eq28}), there are only four 
independent evolution equations: \bea &\disp \delta x^\prime=
&-\left[3\bar{y}^2+3\left(1-\frac{\gamma}{2}\right)\bar{z}^2
+2\bar{u}\bar{x}\right]\delta x+\left(\sqrt{6}\lambda-6\bar{x}-2\bar{u}\right)\bar{y}\delta y\nonumber\\
& &\disp +\left[3\left(\gamma-2\right)\bar{x}-2\bar{u}\right]\bar{z}\delta z
-\left(\bar{x}^2+\bar{y}^2+\bar{z}^2-1\right)\delta u-\delta C_1,\label{eq37}\\
&\disp \delta y^\prime= &-\sqrt{\frac{3}{2}}\lambda\bar{y}\delta x
+3\left[1-3\bar{y}^2+\left(\frac{\gamma}{2}-1\right)\bar{z}^2
-\frac{\lambda}{\sqrt{6}}\bar{x}\right]\delta y
+3\left(\gamma-2\right)\bar{y}\bar{z}\delta z,\label{eq38}\\
&\disp \delta z^\prime= &-6\bar{y}\bar{z}\delta y+3\left[\left(1-\frac{\gamma}{2}\right)-\bar{y}^2
+3\left(\frac{\gamma}{2}-1\right)\bar{z}^2\right]\delta z+\delta C_2,\label{eq39}\\
&\disp \delta u^\prime= &-\bar{u}^2\delta x-2\bar{x}\bar{u}\delta u,\label{eq40}
\eea
where $\delta C_1$ and $\delta C_2$ are the linear perturbations coming from $C_1$ and $C_2$,
respectively. The four eigenvalues of the coefficient matrix of the above equations determine the
stability of the critical point.


\subsection{\label{sect4a} Case~(I)~~$C=0$}

$C=0$ means  {\em no} interaction between hessence and background
matter.  In this case,  one has $C_1=C_2=0$.  It is easy to find
out all critical points
$(\bar{x},\bar{y},\bar{z},\bar{u},\bar{v})$ of the autonomous
system~(\ref{eq32})--(\ref{eq36}). They are required to be real
and satisfy the Friedmann constraint and $\bar{y}\geq 0$,
$\bar{z}\geq 0$. We present them in Table~\ref{tab1}.  Next we
consider the stabilities of these critical points. In this case,
$\delta C_1=\delta C_2=0$. Substituting $\delta C_1$, $\delta C_2$
and the critical point $(\bar{x},\bar{y},\bar{z},\bar{u},\bar{v})$
into Eqs.~(\ref{eq37})--(\ref{eq40}), we find that Points (E.I.1),
(E.I.2), (E.I.3) and (E.I.4) are always unstable, while Point
(E.I.5) and Point (E.I.6) exist and are stable under condition
$\lambda>\sqrt{3\gamma}$ and $\lambda<\sqrt{3\gamma}$,
respectively.

The late time attractor (E.I.5) has
\be{eq41}
\Omega_h=\frac{3\gamma}{\lambda^2},~~~~~~~\Omega_m=1-\frac{3\gamma}{\lambda^2},~~~~~~~
w_h=-1+\gamma,~~~~~~~w_{eff}=-1+\gamma,
\ee
which is a scaling solution. The late time attractor (E.I.6) has
\be{eq42}
\Omega_h=1,~~~~~~~\Omega_m=0,~~~~~~~w_h=-1+\frac{\lambda^2}{3},~~~~~~~
w_{eff}=-1+\frac{\lambda^2}{3},
\ee
which is a hessence-dominated solution. Note that their EoS are all larger than $-1$.

\begin{table}[htbp]
\begin{center}
\begin{tabular}{c|c}
\hline\hline
Label & Critical Point $(\bar{x},\bar{y},\bar{z},\bar{u},\bar{v})$\\ \hline
E.I.1 & $\bar{x}^2\geq 1$,\ 0,\ 0,\ 0,\ $\pm\sqrt{\bar{x}^2-1}$\\
E.I.2 & 0,\ 0,\ 1,\ any,\ 0\\
E.I.3 & $\pm 1$,\ 0,\ 0,\ 0,\ 0\\
E.I.4 & $\frac{\sqrt{6}}{\lambda}$,\ 0,\ 0,\ 0,\ $\pm\sqrt{\frac{6}{\lambda^2}-1}$\\
E.I.5 & $\sqrt{\frac{3}{2}}\frac{\gamma}{\lambda}$,\
     $\sqrt{\frac{3\gamma}{\lambda^2}-\frac{3\gamma^2}{2\lambda^2}}$,\
     $\sqrt{1-\frac{3\gamma}{\lambda^2}}$,\ 0,\ 0\ \\
E.I.6 & $\frac{\lambda}{\sqrt{6}}$,\ $\sqrt{1-\frac{\lambda^2}{6}}$,\ 0,\ 0,\ 0\\ \hline
\end{tabular}
\end{center}
\caption{\label{tab1} Critical points for Case~(I) $C=0$ in the model with exponential potential.}
\end{table}


\subsection{\label{sect4b} Case~(II)~~$C=\alpha\kappa\rho_m\dot{\phi}$}

In this case, $C_1=\sqrt{\frac{3}{2}}\alpha z^2$ and
$C_2=\sqrt{\frac{3}{2}}\alpha xz$.  The physically reasonable
critical points of the autonomous
system~(\ref{eq32})--(\ref{eq36})  are summarized  in
Table~\ref{tab2}.  Next let us consider the stability of these
critical points. In this case, $\delta
C_1=\sqrt{6}\alpha\bar{z}\delta z$ and $\delta
C_2=\sqrt{\frac{3}{2}}\alpha\bar{z}\delta
x+\sqrt{\frac{3}{2}}\alpha\bar{x}\delta z$. Substituting $\delta
C_1$, $\delta C_2$ and the critical point
$(\bar{x},\bar{y},\bar{z},\bar{u},\bar{v})$ into
Eqs.~(\ref{eq37})--(\ref{eq40}), we find that  Point (E.II.1)
 exists and is stable under condition $\alpha<0$ and
$\bar{x}>\max\left\{1,\,\frac{\sqrt{6}}
{\lambda},\,\sqrt{\frac{3}{2}}\frac{\gamma-2}{\alpha}\right\}$;
Point (E.II.2p) exists and is stable under condition
$\alpha<\sqrt{\frac{3}{2}}(\gamma-2)$ and $\lambda>\sqrt{6}$;
Point (E.II.2m) is always unstable; Point (E.II.3) exists and is
stable under condition $\lambda>\sqrt{6\gamma(2-\gamma)}$ and
$\sqrt{\frac{3}{2}}(\gamma-2)<\alpha<\sqrt{\frac{3}{2}}(2-\gamma)$
and $\frac{-\lambda
-\sqrt{\lambda^2-6\gamma(2-\gamma)}}{2}<\alpha<\frac{-\lambda+\sqrt{\lambda^2-6\gamma(2-\gamma)}}{2}$;
Point (E.II.4) exists and is stable under condition
$\lambda<\sqrt{6}$ and
$\alpha<\left(\frac{\gamma}{2}-1\right)\lambda$; Point (E.II.5)
exists and
 is stable under condition
$0>\alpha>\max\left\{\sqrt{\frac{3}{2}}(\gamma-2),\left(\frac{\gamma}{2}-1\right)\lambda\right\}$;
Point (E.II.6) exists and is stable in a suitable
parameter-space~\cite{r38}; and  Point (E.II.7) exists and is
stable under condition $\lambda<\sqrt{6}$ and
$\alpha<\frac{3\gamma}{\lambda}-\lambda$. The corresponding
$\Omega_h$, $\Omega_m$, $w_h$ and $w_{eff}$ of the attractors are
presented in Table~\ref{tab2} as well . Again, we see that their
EoS of hessence and effective EoS are always larger than $-1$
[note that in Point (E.II.6), $\alpha(\lambda+\alpha)+3\gamma>0$ and $\alpha+\lambda>0$ are 
required by the existence of its corresponding $\bar{y}$ and $\bar{z}$, respectively]. 
This implies that the big rip singularity will not appear in this case.

\begin{table}[htbp]
\begin{center}
\begin{tabular}{c|c|cccc}
\hline\hline
Label & Critical Point $(\bar{x},\bar{y},\bar{z},\bar{u},\bar{v})$ & $\Omega_h$ & $\Omega_m$
& $w_h$ & $w_{eff}$\\ \hline
E.II.1 & $\bar{x}^2\geq 1$,\ 0,\ 0,\ 0,\ $\pm\sqrt{\bar{x}^2-1}$  & 1 & 0 & 1 & 1\\
E.II.2p & $+1$,\ 0,\ 0,\ 0,\ 0 & 1 & 0 & 1 & 1\\
E.II.2m & $-1$,\ 0,\ 0,\ 0,\ 0 & --- & --- & --- & ---\\
E.II.3 & $\sqrt{\frac{2}{3}}\frac{\alpha}{\gamma-2}$,\ 0,\ $\sqrt{1-\frac{2\alpha^2}{3(\gamma-2)^2}}$,\ 0,\ 0
            & $\frac{2\alpha^2}{3(\gamma-2)^2}$ & $1-\frac{2\alpha^2}{3(\gamma-2)^2}$ & 1
              & \!\!\!\!\!\!\!$-1+\gamma+\frac{2\alpha^2}{3(2-\gamma)}$\\
E.II.4 & $\frac{\sqrt{6}}{\lambda}$,\ 0,\ 0,\ 0,\ $\pm\sqrt{\frac{6}{\lambda^2}-1}$ & 1 & 0 & 1 & 1\\
E.II.5 & $\sqrt{\frac{3}{2}}\frac{\gamma-2}{\alpha}$,\ 0,\ 0,\ 0,\ $\pm\sqrt{\frac{3(\gamma-2)^2}{2\alpha^2}-1}$
            & 1 & 0 & 1 & 1\\
E.II.6 & $\sqrt{\frac{3}{2}}\frac{\gamma}{\lambda+\alpha}$,\
       $\sqrt{\frac{2\alpha^2-3(\gamma-2)\gamma+2\alpha\lambda}{2(\alpha+\lambda)^2}}$,\
       $\sqrt{\frac{\lambda(\lambda+\alpha)-3\gamma}{(\lambda+\alpha)^2}}$,\ 0,\ 0\,
            & \,$\frac{\alpha(\lambda+\alpha)+3\gamma}{(\lambda+\alpha)^2}$
            & $\frac{\lambda(\lambda+\alpha)-3\gamma}{(\lambda+\alpha)^2}$
            & $-1+\frac{3\gamma^2}{\alpha(\lambda+\alpha)+3\gamma}$
             & $-1+\frac{\gamma\lambda}{\alpha+\lambda}$\\
E.II.7 & $\frac{\lambda}{\sqrt{6}}$,\ $\sqrt{1-\frac{\lambda^2}{6}}$,\ 0,\ 0,\ 0
            & 1 & 0 & $-1+\frac{\lambda^2}{3}$ & $-1+\frac{\lambda^2}{3}$\\ \hline
\end{tabular}
\end{center}
\caption{\label{tab2} Critical points for Case~(II) $C=\alpha\kappa\rho_m\dot{\phi}$ in the model
with exponential potential.}
\end{table}


\subsection{\label{sect4c} Case~(III)~~$C=3\beta H\rho_{tot}=3\beta H\left(\rho_h+\rho_m\right)$}

In this case, $C_1=\frac{3}{2}\beta x^{-1}$ and
$C_2=\frac{3}{2}\beta z^{-1}$. The physically reasonable critical
points of the autonomous system~(\ref{eq32})--(\ref{eq36}) are
presented in Table~\ref{tab3}, where
\be{eq43}
r_3\equiv\sqrt{1+\frac{4\beta}{2-\gamma}}.
  \ee
   Points (E.III.4),
(E.III.5) and (E.III.6) are the three real solutions of \bea
&&\frac{2}{\gamma}\bar{x}\left(\frac{\lambda}{\sqrt{6}}-\bar{x}\right)
\left(\gamma-\sqrt{\frac{2}{3}}\lambda\bar{x}\right)=\beta,\label{eq44}\\
&&\bar{y}^2=\left(\frac{2}{\gamma}-1\right)\bar{x}^2
-\sqrt{\frac{2}{3}}\frac{\lambda}{\gamma}\bar{x}+1,\label{eq45}\\
&&\bar{z}^2=\frac{2}{\gamma}\left(-\bar{x}^2+\frac{\lambda}{\sqrt{6}}\bar{x}\right),\label{eq46}
\eea
with $\bar{u}=\bar{v}=0$ and requiring $\bar{y}\geq 0$, $\bar{z}\geq 0$ by definition. 
Although the real roots of the Eqs.~(\ref{eq44})--(\ref{eq46}) 
are easy to obtain, we do not present them here, since they are long in length. 
Note that Point (E.III.3) exists only when $\beta>0$ and $\gamma>2$, which 
is beyond the range $0<\gamma\leq 2$.

Next, we consider the stabilities of these critical points. In
this case, $\delta C_1=-\frac{3}{2}\beta\bar{x}^{-2}\delta x$ and
$\delta C_2=-\frac{3}{2}\beta\bar{z}^{-2}\delta z$. Substituting
$\delta C_1$, $\delta C_2$ and the critical point
$(\bar{x},\bar{y},\bar{z},\bar{u},\bar{v})$ into
Eqs.~(\ref{eq37})--(\ref{eq40}), we see that Points (E.III.1p),
(E.III.1m) and (E.III.2m) are always unstable, while Point
(E.III.2p) exists and is stable under condition
$0>\beta>(\gamma-2)/4$ and
$\lambda>\left[12+3(1+r_3)(-2+\gamma)\right]/\left(2\sqrt{3}\sqrt{1-r_3}\right)$.
Points (E.III.4), (E.III.5) and (E.III.6) exist and are stable in
proper parameter-space~\cite{r38}.

The late time attractor (E.III.2p) has \be{eq47}
\Omega_h=\frac{1}{2}\left(1-r_3\right),~~~~~~~
\Omega_m=\frac{1}{2}\left(1+r_3\right),~~~~~~~
w_h=1,~~~~~~~w_{eff}=1+\frac{1}{2}\left(\gamma-2\right)\left(1+r_3\right),
\ee which is a scaling solution. The late time attractors
(E.III.4), (E.III.5) and (E.III.6) have \be{eq48}
\Omega_h=\bar{x}^2+\bar{y}^2,~~~~~~~\Omega_m=\bar{z}^2,~~~~~~~
w_h=-1+\frac{2\bar{x}^2}{\bar{x}^2+\bar{y}^2},~~~~~~~
w_{eff}=-1+\sqrt{\frac{2}{3}}\lambda\bar{x}.
  \ee
They are all scaling solutions. It is easy to see that their EoS
of hessence and effective EoS are all larger than $-1$ [note that
$0<\gamma\leq 2$ and $0<r_3<1$ for Point (E.III.2p), while
$\bar{x}\geq 0$ is required by Eq.~(\ref{eq46}) for Points
(E.III.4), (E.III.5) and (E.III.6)].

\begin{table}[htbp]
\begin{center}
\begin{tabular}{ccc}
\hline\hline
Label &\vline & Critical Point $(\bar{x},\bar{y},\bar{z},\bar{u},\bar{v})$ \\ \hline
E.III.1p &\vline & $\left[\frac{1}{2}\left(1+r_3\right)\right]^{1/2}$,\ 0,\
         $\left[\frac{1}{2}\left(1-r_3\right)\right]^{1/2}$,\ 0,\ 0\\
E.III.1m &\vline & $-\left[\frac{1}{2}\left(1+r_3\right)\right]^{1/2}$,\ 0,\
         $\left[\frac{1}{2}\left(1-r_3\right)\right]^{1/2}$,\ 0,\ 0\\
E.III.2p &\vline & $\left[\frac{1}{2}\left(1-r_3\right)\right]^{1/2}$,\ 0,\
         $\left[\frac{1}{2}\left(1+r_3\right)\right]^{1/2}$,\ 0,\ 0\\
E.III.2m &\vline & $-\left[\frac{1}{2}\left(1-r_3\right)\right]^{1/2}$,\ 0,\
         $\left[\frac{1}{2}\left(1+r_3\right)\right]^{1/2}$,\ 0,\ 0\\
E.III.3 &\vline & $\frac{\sqrt{6}}{\lambda}$,\ $\sqrt{\frac{\beta}{2}}$,\
        $\sqrt{\frac{\beta}{-2+\gamma}}$,\ 0,\
        $\pm\sqrt{-1+\frac{\beta\gamma}{-4+2\gamma}+\frac{6}{\lambda^2}}$\\ \hline
Note: & &For Points (E.III.4), (E.III.5) and (E.III.6), see text\\ \hline
\end{tabular}
\end{center}
\caption{\label{tab3} Critical points for Case~(III) $C=3\beta
H\rho_{tot}=3\beta H\left(\rho_h+\rho_m\right)$ in the model with
exponential potential. $r_3$ is given in Eq.~(\ref{eq43}).}
\end{table}


\subsection{\label{sect4d} Case~(IV)~~$C=3\eta H\rho_m$}

In this case, $C_1=\frac{3}{2}\eta x^{-1}z^2$ and
$C_2=\frac{3}{2}\eta z$.  We show the physically reasonable
critical points of the autonomous
system~(\ref{eq32})--(\ref{eq36}) in Table~\ref{tab4}, where
\be{eq49}
r_y\equiv\sqrt{\frac{-3\gamma^3+6\gamma^2(1+\eta)-3\gamma\eta(4+\eta)+2\eta(3\eta+\lambda^2)}{2\gamma\lambda^2}},
\ee \be{eq50}
r_z\equiv\sqrt{\frac{(\gamma-\eta)\lambda^2-3(\gamma-\eta)^2}{\gamma\lambda^2}}.
\ee
 To study the stability of these critical points, we obtain 
$\delta C_1=-\frac{3}{2}\eta\bar{x}^{-2}\bar{z}^2\delta
x+3\eta\bar{x}^{-1}\bar{z}\delta z$ and $\delta
C_2=\frac{3}{2}\eta\delta z$ by linearizing $C_1$ and $C_2$. 
Substituting $\delta C_1$, $\delta C_2$ and the critical point 
$(\bar{x},\bar{y},\bar{z},\bar{u},\bar{v})$ into
Eqs.~(\ref{eq37})--(\ref{eq40}), we find that Point (E.IV.1)
exists and is stable under condition $\eta<-2+\gamma$ and
$\bar{x}>\max\left\{1,\frac{\sqrt{6}}{\lambda}\right\}$; Point
(E.IV.2p) exists and is stable under condition $0>\eta>\gamma-2$
and
$\lambda>\sqrt{\frac{3}{2}}(\gamma-\eta)\sqrt{\frac{\gamma-2}{\eta}}$;
Point (E.IV.2m) is always unstable; Point (E.IV.3p) exists and is
stable under condition $\eta<\gamma-2$ and $\lambda>\sqrt{6}$;
Point (E.IV.3m) is always unstable; Point (E.IV.4) exists and is
stable under condition $\eta<\gamma-2$ and $\lambda<\sqrt{6}$;
Point (E.IV.5) exists and is stable under condition $\eta<\gamma$
and $\lambda<\min\left\{\sqrt{6},\sqrt{3(\gamma-\eta)}\right\}$;
and  Point (E.IV.6) exists and is stable in proper
parameter-space~\cite{r38}.

The corresponding $\Omega_h$, $\Omega_m$, $w_h$ and $w_{eff}$ of
the attractors are presented in Table~\ref{tab4} as well. Once
again, we see that their EoS of hessence and effective EoS are all
larger than $-1$ [note that $\eta<0$ is required by the existence
of its corresponding $\bar{x}$ for Point (E.IV.2p), while
$\gamma-\eta>0$ is required by Eq.~(\ref{eq50}) for Point
(E.IV.6)].

\begin{table}[htbp]
\begin{center}
\begin{tabular}{c|c|cccc}
\hline\hline
Label & Critical Point $(\bar{x},\bar{y},\bar{z},\bar{u},\bar{v})$ & $\Omega_h$ & $\Omega_m$
& $w_h$ & $w_{eff}$\\ \hline
E.IV.1 & $\bar{x}^2\geq 1$,\ 0,\ 0,\ 0,\ $\pm\sqrt{\bar{x}^2-1}$  & 1 & 0 & 1 & 1\\
E.IV.2p & $\sqrt{\frac{\eta}{\gamma-2}}$,\ 0,\ $\sqrt{1-\frac{\eta}{\gamma-2}}$,\ 0,\ 0
          & $\frac{\eta}{\gamma-2}$ & $1-\frac{\eta}{\gamma-2}$ & 1
           & $-1+\gamma-\eta$\\
E.IV.2m & \,$-\sqrt{\frac{\eta}{\gamma-2}}$,\ 0,\ $\sqrt{1-\frac{\eta}{\gamma-2}}$,\ 0,\ 0\,
            & --- & --- & --- & ---\\
E.IV.3p & $+1$,\ 0,\ 0,\ 0,\ 0 & 1 & 0 & 1 & 1\\
E.IV.3m & $-1$,\ 0,\ 0,\ 0,\ 0 & --- & --- & --- & ---\\
E.IV.4 & $\frac{\sqrt{6}}{\lambda}$,\ 0,\ 0,\ 0,\ $\pm\sqrt{\frac{6}{\lambda^2}-1}$ & 1 & 0 & 1 & 1\\
E.IV.5 & $\frac{\lambda}{\sqrt{6}}$,\ $\sqrt{1-\frac{\lambda^2}{6}}$,\ 0,\ 0,\ 0
            & 1 & 0 & $-1+\frac{\lambda^2}{3}$ & $-1+\frac{\lambda^2}{3}$\\
E.IV.6 & $\sqrt{\frac{3}{2}}\frac{\gamma-\eta}{\lambda}$,\ $r_y$,\ $r_z$,\ 0,\ 0
            & $1-r_{z}^2$ & $r_{z}^2$  & $-1+\frac{3(\gamma-\eta)^2}{\lambda^2(1-r_{z}^2)}$
             & \,\,$-1+\gamma-\eta$\\ \hline
\end{tabular}
\end{center}
\caption{\label{tab4} Critical points for Case~(IV) $C=3\eta
H\rho_m$ in the model with exponential potential. $r_y$ and $r_z$
are given in Eqs.~(\ref{eq49}) and (\ref{eq50}), respectively.}
\end{table}


\section{\label{sect5} Model with (inverse) power law potential}

In this section, we consider the hessence model with a (inverse)
power law potential
 \be{eq51}
V(\phi)=V_0\left(\kappa\phi\right)^n,
 \ee
 where $n$ is a
dimensionless constant. $V(\phi)$ is a power law potential when
$n>0$ while it is an inverse power law potential when $n<0$. In
this case, Eqs.~(\ref{eq21})--(\ref{eq25}) become \bea &&\disp
x^\prime=3x\left(x^2-v^2+\frac{\gamma}{2}z^2-1\right)-uv^2
-\frac{n}{2}uy^2-C_1,\label{eq52}\\
&&\disp y^\prime=3y\left(x^2-v^2+\frac{\gamma}{2}z^2\right)
+\frac{n}{2}uxy,\label{eq53}\\
&&\disp z^\prime=3z\left(x^2-v^2+\frac{\gamma}{2}z^2-\frac{\gamma}{2}\right)+C_2,\label{eq54}\\
&&\disp u^\prime=-xu^2,\label{eq55}\\
&&\disp
v^\prime=3v\left(x^2-v^2+\frac{\gamma}{2}z^2-1\right)-xuv,\label{eq56}
\eea where $C_1$ and $C_2$ are defined in Eq.~(\ref{eq26}) and
depend on interaction form $C$. In the following subsections, we
will consider four interaction forms $C$ given in the end of
Sec.~\ref{sect3}.

To study the stability of the critical points of
Eqs.~(\ref{eq52})--(\ref{eq56}), we substitute linear
perturbations $x\to\bar{x}+\delta x$, $y\to\bar{y}+\delta y$,
$z\to\bar{z}+\delta z$, $u\to\bar{u}+\delta u$, and
$v\to\bar{v}+\delta v$ about the critical point
$(\bar{x},\bar{y},\bar{z},\bar{u},\bar{v})$ into
Eqs.~(\ref{eq52})--(\ref{eq56}) and linearize them. Note that
these critical points must satisfy the Friedmann constraint,
$\bar{y}\geq 0$, $\bar{z}\geq 0$ and requirement of
$\bar{x},\,\bar{y},\,\bar{z},\,\bar{u},\,\bar{v}$ all being real.
Because of the Friedmann constraint~(\ref{eq28}), there are only four 
independent evolution equations: \bea &\disp \delta x^\prime=
&\left\{3\left[\left(\frac{\gamma}{2}-1\right)\bar{z}^2
-\bar{y}^2\right]-2\bar{u}\bar{x}\right\}\delta x
-\left[6\bar{x}\bar{y}+\left(2+n\right)\bar{u}\bar{y}\right]\delta
y
+\left[3\left(\gamma-2\right)\bar{x}-2\bar{u}\right]\bar{z}\delta z\nonumber\\
& &\disp -\left[\bar{x}^2+\left(1+\frac{n}{2}\right)\bar{y}^2+\bar{z}^2-1\right]\delta u
-\delta C_1,\label{eq57}\\
&\disp \delta y^\prime= &\frac{n}{2}\bar{y}\bar{u}\delta x+3\left[1-3\bar{y}^2+
\left(\frac{\gamma}{2}-1\right)\bar{z}^2+\frac{n}{6}\bar{u}\bar{x}\right]\delta y
+3\left(\gamma-2\right)\bar{y}\bar{z}\delta z+\frac{n}{2}\bar{x}\bar{y}\delta u,\label{eq58}\\
&\disp \delta z^\prime= &-6\bar{y}\bar{z}\delta y+\left\{\left(\gamma-2\right)\bar{z}
+3\left[\left(1-\frac{\gamma}{2}\right)-\bar{y}^2+\left(\frac{\gamma}{2}
-1\right)\bar{z}^2\right]\right\}\delta z+\delta C_2,\label{eq59}\\
&\disp \delta u^\prime= &-\bar{u}^2\delta x-2\bar{x}\bar{u}\delta
u,\label{eq60}
\eea
 where $\delta C_1$ and $\delta C_2$ are the
linear perturbations coming from $C_1$ and $C_2$, respectively.
Again, the four eigenvalues of the coefficient matrix of the above
equations determine the stability of the critical points.


\subsection{\label{sect5a} Case~(I)~~$C=0$}
In this case, $C_1=C_2=0$, namely there is {\em no} interaction
between hessence and background matter. It is easy to find out all
critical points $(\bar{x},\bar{y},\bar{z},\bar{u},\bar{v})$ of the
autonomous system~(\ref{eq52})--(\ref{eq56}). We present them in
Table~\ref{tab5}. Checking the eigenvalues of
Eqs.(\ref{eq57})-(\ref{eq60}), we find that Points (P.I.1),
(P.I.2), and (P.I.4) are always unstable, while Point (P.I.3) is
always stable. The unique late time attractor (P.I.3) has
\be{eq61}
\Omega_h=1,~~~~~~~\Omega_m=0,~~~~~~~w_h=-1,~~~~~~~w_{eff}=-1, \ee
which is a hessence-dominated solution and it is an asymptotical
de Sitter attractor.

\begin{table}[htbp]
\begin{center}
\begin{tabular}{c|c}
\hline\hline
Label & Critical Point $(\bar{x},\bar{y},\bar{z},\bar{u},\bar{v})$\\ \hline
P.I.1 & $\bar{x}^2\geq 1$,\ 0,\ 0,\ 0,\ $\pm\sqrt{\bar{x}^2-1}$\\
P.I.2 & 0,\ 0,\ 1,\ any,\ 0\\
P.I.3 & 0,\ 1,\ 0,\ 0,\ 0\\
P.I.4 & $\pm 1$,\ 0,\ 0,\ 0,\ 0\\
\hline
\end{tabular}
\end{center}
\caption{\label{tab5} Critical points for Case~(I) $C=0$ in the model
with (inverse) power law potential.}
\end{table}


\subsection{\label{sect5b} Case~(II)~~$C=\alpha\kappa\rho_m\dot{\phi}$}

In this case, $C_1=\sqrt{\frac{3}{2}}\alpha z^2$ and
$C_2=\sqrt{\frac{3}{2}}\alpha xz$. We summarize the physically
reasonable critical points of the autonomous
system~(\ref{eq52})--(\ref{eq56}) in Table~\ref{tab6}. Next, we
consider the stabilities of these critical points. In this case,
$\delta C_1=\sqrt{6}\alpha\bar{z}\delta z$ and $\delta
C_2=\sqrt{\frac{3}{2}}\alpha\bar{z}\delta
x+\sqrt{\frac{3}{2}}\alpha\bar{x}\delta z$.
 The four eigenvalues of the
coefficient matrix of Eqs.~(\ref{eq57})--(\ref{eq60}) tell us that
 Points (P.II.1), (P.II.2), (P.II.3) and (P.II.4) are always
unstable, while Point (P.II.5) is always stable. The unique late
time attractor (P.II.5) has \be{eq62}
\Omega_h=1,~~~~~~~\Omega_m=0,~~~~~~~w_h=-1,~~~~~~~w_{eff}=-1, \ee
which is a hessence-dominated solution and it is an asymptotical
de Sitter attractor.

\begin{table}[htbp]
\begin{center}
\begin{tabular}{c|c}
\hline\hline
Label & Critical Point $(\bar{x},\bar{y},\bar{z},\bar{u},\bar{v})$\\ \hline
P.II.1 & $\bar{x}^2\geq 1$,\ 0,\ 0,\ 0,\ $\pm\sqrt{\bar{x}^2-1}$\\
P.II.2 & $\pm 1$,\ 0,\ 0,\ 0,\ 0\\
P.II.3 & $\sqrt{\frac{2}{3}}\frac{\alpha}{\gamma-2}$,\ 0,\
        $\sqrt{1-\frac{2\alpha^2}{3(\gamma-2)^2}}$,\ 0,\ 0\\
P.II.4 & $\sqrt{\frac{3}{2}}\frac{\gamma-2}{\alpha}$,\ 0,\ 0,\ 0,\
        $\pm\sqrt{-1+\frac{3(\gamma-2)^2}{2\alpha^2}}$\\
P.II.5 & 0,\ 1,\ 0,\ 0,\ 0\\\hline
\end{tabular}
\end{center}
\caption{\label{tab6} Critical points for Case~(II) $C=\alpha\kappa\rho_m\dot{\phi}$ in the model
with (inverse) power law potential.}
\end{table}


\subsection{\label{sect5c} Case~(III)~~$C=3\beta H\rho_{tot}=3\beta H\left(\rho_h+\rho_m\right)$}

In this case, $C_1=\frac{3}{2}\beta x^{-1}$ and
$C_2=\frac{3}{2}\beta z^{-1}$. The physically reasonable critical
points of the autonomous system~(\ref{eq52})--(\ref{eq56}) are
summarized in Table~\ref{tab7}. There, $r_3$ is defined by
Eq.~(\ref{eq43}). Substituting $\delta
C_1=-\frac{3}{2}\beta\bar{x}^{-2}\delta x$, $\delta
C_2=-\frac{3}{2}\beta\bar{z}^{-2}\delta z$, and  the critical
point $(\bar{x},\bar{y},\bar{z},\bar{u},\bar{v})$ into
Eqs.~(\ref{eq57})--(\ref{eq60}), we find that {\em no} stable
attractor exists in this case.

\begin{table}[htbp]
\begin{center}
\begin{tabular}{c|c}
\hline\hline
Label & Critical Point $(\bar{x},\bar{y},\bar{z},\bar{u},\bar{v})$ \\ \hline
P.III.1p & $\left[\frac{1}{2}\left(1+r_3\right)\right]^{1/2}$,\ 0,\
         $\left[\frac{1}{2}\left(1-r_3\right)\right]^{1/2}$,\ 0,\ 0\\
P.III.1m & \,$-\left[\frac{1}{2}\left(1+r_3\right)\right]^{1/2}$,\ 0,\
         $\left[\frac{1}{2}\left(1-r_3\right)\right]^{1/2}$,\ 0,\ 0\\
P.III.2p & $\left[\frac{1}{2}\left(1-r_3\right)\right]^{1/2}$,\ 0,\
         $\left[\frac{1}{2}\left(1+r_3\right)\right]^{1/2}$,\ 0,\ 0\\
P.III.2m & \,$-\left[\frac{1}{2}\left(1-r_3\right)\right]^{1/2}$,\ 0,\
         $\left[\frac{1}{2}\left(1+r_3\right)\right]^{1/2}$,\ 0,\ 0\\ \hline
\end{tabular}
\end{center}
\caption{\label{tab7} Critical points for Case~(III) $C=3\beta
H\rho_{tot}=3\beta H\left(\rho_h+\rho_m\right)$ in the model with
(inverse) power law potential. $r_3$ is given in
Eq.~(\ref{eq43}).}
\end{table}


\subsection{\label{sect5d} Case~(IV)~~$C=3\eta H\rho_m$}

In this case, $C_1=\frac{3}{2}\eta x^{-1}z^2$ and
$C_2=\frac{3}{2}\eta z$, we have the physically reasonable
critical points shown in Table~\ref{tab8}. Substituting $\delta
C_1=-\frac{3}{2}\eta\bar{x}^{-2}\bar{z}^2\delta
x+3\eta\bar{x}^{-1}\bar{z}\delta z$, $\delta
C_2=\frac{3}{2}\eta\delta z$, and the critical point
$(\bar{x},\bar{y},\bar{z},\bar{u},\bar{v})$ into
Eqs.~(\ref{eq57})--(\ref{eq60}), the four eigenvalues of the
coefficient matrix of the resulting equations tell us that {\em
no} stable attractor exists as well in this case.

\begin{table}[htbp]
\begin{center}
\begin{tabular}{c|c}
\hline\hline
Label & Critical Point $(\bar{x},\bar{y},\bar{z},\bar{u},\bar{v})$ \\ \hline
P.IV.1 & $\bar{x}^2\geq 1$,\ 0,\ 0,\ 0,\ $\pm\sqrt{\bar{x}^2-1}$\\
P.IV.2 & $\pm\sqrt{\frac{\eta}{\gamma-2}}$,\ 0,\ $\sqrt{1-\frac{\eta}{\gamma-2}}$,\ 0,\ 0\\
P.IV.3 & $\pm 1$,\ 0,\ 0,\ 0,\ 0\\
\hline
\end{tabular}
\end{center}
\caption{\label{tab8} Critical points for Case~(IV) $C=3\eta H\rho_m$ in the model
with (inverse) power law potential.}
\end{table}


\section{\label{sect6} No big rip in hessence model}

Obviously, we can see from Sec.~\ref{sect4} and Sec.~\ref{sect5}
that $w_h$ and $w_{eff}$ of all stable late time attractors are
larger than or equal to $-1$ for these two particular models with exponential
or (inverse) power law potential and the interaction term $C$ is
chosen to be four different forms. {\em No} phantom-like late time
attractor with $w_h$ or $w_{eff}$ less than $-1$ can exist in the
hessence model of dark energy. However, one may wonder whether
this observation depends on the forms of potential $V(\phi)$ and 
the interaction term $C$. Therefore, it is interesting to
investigate this feature in a general case.

Let us come back to the most general equations system, i.e.
Eqs.~(\ref{eq21})--(\ref{eq25}). Now, we leave $V(\phi)$ and $C$
undetermined, except for assuming that they can make the equations 
system closed, in other words, Eqs.~(\ref{eq21})--(\ref{eq25}) is
an autonomous system. The critical point
$(\bar{x},\bar{y},\bar{z},\bar{u},\bar{v})$ of the autonomous
system satisfies
$\bar{x}^\prime=\bar{y}^\prime=\bar{z}^\prime=\bar{u}^\prime=\bar{v}^\prime=0$.
Of course, they are also subject to the Friedmann constraint, i.e.
$\bar{x}^2+\bar{y}^2+\bar{z}^2-\bar{v}^2=1$. From
Eq.~(\ref{eq24}), $\bar{u}^\prime=0$, we have $\bar{x}\bar{u}=0$.
 Substituting this into Eq.~(\ref{eq25}),
$\bar{v}^\prime=0$, we have either $\bar{v}=0$ or
$\bar{x}^2-\bar{v}^2+\frac{\gamma}{2}\bar{z}^2-1=0$. If the latter
holds, from Eq.~(\ref{eq29}), we have
$\Omega_m=\bar{z}^2=\frac{2}{\gamma}\left[1-\left(\bar{x}^2
-\bar{v}^2\right)\right]$.  Physics requires  $\Omega_m\leq 1$.
Therefore $\bar{x}^2-\bar{v}^2\geq 0$ is required for the case
$0<\gamma\leq 2$.  We see from Eq.~(\ref{eq30}) that $w_h\geq -1$
in this case. If
$\bar{x}^2-\bar{v}^2+\frac{\gamma}{2}\bar{z}^2-1\not=0$, we then
have $\bar{v}=0$ as mentioned above. Obviously, from
Eq.~(\ref{eq30}), $w_h\geq -1$ is inevitable. On the other hand,
we can see no big rip in the hessence model as follows. Since
$\bar{x}^2-\bar{v}^2\geq 0$ for both cases mentioned above, we can
see from Eq.~(\ref{eq27}) that $\dot{H}\leq 0$, which implies
$w_{eff}\geq -1$, cf. Eq.~(\ref{eq16}). Therefore, we conclude
that $w_h\geq -1$ and $w_{eff}\geq -1$ always hold for all
critical points of the autonomous system
(\ref{eq21})--(\ref{eq25}). {\em No} phantom-like late time
attractor with EoS less than $-1$ can exist. This result is
independent of the form of potential $V(\phi)$ and interaction
form $C$.

Therefore,  EoS less than $-1$ is transient in the hessence model.
Eventually, it will go to quintessence-like attractors whose EoS
is larger than $-1$ or asymptotically to de Sitter attractor whose
EoS is a constant $-1$. Thus, the big rip will not appear in the
hessence model.


\section{\label{sect7} Conclusion and discussions}
In this work, the cosmological evolution of hessence dark energy
is investigated. We considered two models with exponential and
(inverse) power law potentials of hessence respectively, and
investigated the dynamical system for the four cases with
different interactions between hessence and background perfect
fluid. By the phase-space analysis, we find that some stable
attractors can exist, which are either scaling solutions or
hessence-dominated solutions with EoS larger than or equal to
$-1$. {\em No} phantom-like late time attractors with EoS less
than $-1$ can exist. We have shown that this essential result
still holds in a general case beyond particular potentials 
and interaction forms. Thus, the big rip will not appear in the
hessence model.

Our result of the hessence model is very different from that of the 
quintom model studied in~\cite{r22,r23}, where the
phantom-dominated solution is the unique late time attractor and
the big rip is inevitable. The difference between our result 
and that of \cite{r22,r23} is not due to the interaction between
the hessence and background matter, since our conclusion still
holds for the case $C=0$, that is, the case without the
interaction. The difference should be due to the form of the
interaction between $\phi_1$ and $\phi_2$ [cf.
Eqs.~(\ref{eq4})--(\ref{eq7})]. Guo {\it et al}~\cite{r22} and
Zhang {\it et al}~\cite{r23} only studied the cosmological
evolution of the quintom model without direct coupling between
$\phi_1$ and $\phi_2$, i.e.
$V(\phi_1,\phi_2)=V_{\phi_1}+V_{\phi_2} \equiv
V_{\phi_{1}0}\exp(-\lambda_{\phi_1}\kappa\phi_1)
+V_{\phi_{2}0}\exp(-\lambda_{\phi_2}\kappa\phi_2),$ and with a
special interaction between $\phi_1$ and $\phi_2$, i.e.
$V(\phi_1,\phi_2)=V_{\phi_1}+V_{\phi_2}+V_{int}$ and
$V_{int}\sim\left(V_{\phi_1}V_{\phi_2}\right)^{1/2}$,
respectively. As pointed out in~\cite{r28}, in the hessence model,
the potential is imposed to be the form of $V(\phi)$, or
equivalently, $V(\phi_1^2-\phi_2^2)$ in terms of $\phi_1$ and
$\phi_2$ [cf. Eqs.~(\ref{eq4})--(\ref{eq7})]. Except for the very
special case with $V(\phi)\sim\phi^2$, the two fields $\phi_1$ and
$\phi_2$ are coupled in the general case, and the interaction between 
$\phi_1$ and $\phi_2$ is quite complicated, rather than the very special 
interaction considered in~\cite{r23}.

Another issue is about the fate of our universe. From our result,
the present super-acceleration of universe, i.e. $\dot{H}>0$, is
transient and our universe can avoid the fate of big rip. It can
either accelerate forever ($w_{eff}<-1/3$) or come back to
deceleration ($w_{eff}>-1/3$), which is determined by the model
parameters.


\section*{ACKNOWLEDGMENTS}
HW is grateful to Zong-Kuan Guo, Yun-Song Piao, Bo Feng, Li-Ming Cao, Da-Wei Pang, 
Yi Zhang, Hui Li, Qi Guo, Hong-Sheng Zhang, Ding-Fang Zeng, Xin Zhang, Ling-Mei Cheng 
and Fei Wang for helpful discussions. This work was supported in part by a grant from Chinese 
Academy of Sciences, a grant from NSFC, China (No. 10325525 and No. 90403029), 
and a grant from the Ministry of Science and Technology of China (No. TG1999075401).

\renewcommand{\baselinestretch}{1.2}



\begin{thebibliography}{99}
\bibitem{r38}
In fact, the particular parameter-space for the existence and stability of this critical point is considerably involved
and verbose. Since our main aim is to point out the fact that it can exist and is stable, the corresponding
parameter-space is not necessary to be presented here. One can work it out comfortably by the help of computer,
esp. by the help of Mathematica.
\end{thebibliography}
\end{document}